\documentclass{article}

\usepackage{graphicx}
\usepackage{amsmath}
\usepackage{amssymb}
\usepackage{mathtools}
\usepackage{amsthm}
\usepackage[textsize=tiny]{todonotes}
\usepackage{subcaption} 
\usepackage{multirow}
\usepackage{arydshln} 
\usepackage{booktabs}
\usepackage{tabularx}
\usepackage{array}
\usepackage{makecell}
\usepackage{threeparttable}
\usepackage{longtable}
\usepackage{colortbl} 
\usepackage[acronym]{glossaries}
\usepackage{caption} 
\usepackage{wrapfig}
\usepackage{adjustbox} 
\usepackage{pifont} 
\usepackage{xcolor}

\newcommand{\cmark}{\textcolor{green!60!black}{\ding{51}}} 
\newcommand{\xmark}{\textcolor{red}{\ding{55}}}            
\newcommand{\dash}{$\circ$} 

\usepackage[T1]{fontenc}
\usepackage[shortlabels]{enumitem}
\setlist[enumerate]{nosep}

\makeglossaries

\newacronym{MAS}{MAS}{Multi-Agent Systems}
\newacronym{LLM}{LLM}{Large Language Models}
\newacronym{AI}{AI}{Artificial Intelligence}
\newacronym{RAG}{RAG}{Retrieval-Augmented Generation}
\usepackage[preprint]{neurips_2025}

\usepackage[utf8]{inputenc} 
\usepackage[T1]{fontenc}    
\usepackage{hyperref}       
\usepackage{url}            
\usepackage{booktabs}       
\usepackage{amsfonts}       
\usepackage{nicefrac}       
\usepackage{microtype}      
\usepackage{xcolor}         

\title{Stop Reducing Responsibility in LLM-Powered Multi-Agent Systems to Local Alignment}

\author{%
  Jinwei Hu$^{1}$,
  Yi Dong$^{1}$,
  Shuang Ao$^{2}$,
  Zhuoyun Li$^{1}$, 
  Boxuan Wang$^{1}$,
  Lokesh Singh$^{2}$, \And
  Guangliang Cheng$^{1}$,
  Sarvapali D. Ramchurn$^{2}$,
  Xiaowei Huang$^{1}$ \\
  \\
  $^{1}$School of Computer Science and Informatics, University of Liverpool, UK \\
  $^{2}$Department of Electronics and Computer Science, University of Southampton, UK \\
}

\begin{document}

\maketitle

\begin{abstract}
  LLM-powered Multi-Agent Systems (LLM-MAS) unlock new potentials in distributed reasoning, collaboration, and task generalization but also introduce additional risks due to unguaranteed agreement, cascading uncertainty, and adversarial vulnerabilities. We argue that ensuring responsible behavior in such systems requires a paradigm shift: \textbf{\textit{from local, superficial agent-level alignment to global, systemic agreement.}} We conceptualize responsibility not as a static constraint but as a lifecycle-wide property encompassing \textit{agreement}, \textit{uncertainty}, and \textit{security}, each requiring the complementary integration of \textit{subjective human-centered values and objective verifiability}. Furthermore, a \textit{dual-perspective governance} framework that combines interdisciplinary design with human-AI collaborative oversight is essential for tracing and ensuring responsibility throughout the lifecycle of LLM-MAS. Our position views LLM-MAS not as loose collections of agents, but as unified, dynamic socio-technical systems that demand principled mechanisms to support each dimension of responsibility and enable ethically aligned, verifiably coherent, and resilient behavior for sustained, system-wide agreement.
\end{abstract}

\section{Introduction}
\gls{MAS} constitute a foundational research domain in decision-making, where autonomous agents\footnote{$Agent$ could be software agent, robotics agent, embodied agent or human agent.} interact to pursue individual or collective goals within dynamic environments \cite{hu2025tapasfreetrainingfreeadaptation}. The recent emergence of \gls{LLM} has profoundly transformed \gls{MAS} by introducing agents capable of sophisticated natural language understanding, extensive knowledge representation, and advanced reasoning capabilities \cite{bubeck2023sparksartificialgeneralintelligence}. Techniques such as LangChain \cite{topsakal2023creating} and Retrieval-Augmented Generation (RAG) \cite{lewis2020retrieval} have further enhanced these capabilities, enabling LLM-empowed agents to dynamically integrate external knowledge and tools, thus significantly amplifying the scale and efficiency of collective problem-solving beyond human limitations.

However, the integration of \gls{LLM} into \gls{MAS} introduces new challenges. 
Unlike traditional \gls{MAS} that rely on predefined protocols and \textit{deterministic behaviors}, LLM-based agents trained on vast  datasets may exhibit emergent, context-dependent, and inherently \textit{unpredictable behaviors} \cite{cheng2024exploring}. Such unpredictability necessitates more comprehensive framework to ensure reliable decision-making and collective convergence, fundamentally reshaping the local LLM responsibility in intelligent systems to global systemic responsibility, which refers to \textbf{\textit{a lifecycle-wide capacity to maintain continuous agreement with human values while achieving system goals, supported by quantifiable, verifiable, and traceable metrics that enable dynamic evaluation and safe control of agent behaviors.}}

Contemporary LLM-MAS applications often exhibit a critical gap in responsibility management. We summarize ten representative categories and their corresponding applications in Table~\ref{tab:llmmas_responsibility} based on~\cite{li2024survey,guo2024large}, showing that LLMs are frequently embedded or assigned diverse roles based solely on local alignment, with limited attention to multiple responsible considerations spanning agreement (to enable coherent agent states), uncertainty (to handle internal and external stochasticity), security (to ensure resilience against adversarial behaviors), and governance (to support adaptability and stability of oversight). This narrow focus risks systemic inconsistency as agent populations scale or interactions evolve, undermining the dependability of LLM-based multi-agent systems in real-world scenarios \cite{nascimento2023self, han2024llm}.

In this paper, we advocate all aspects of lifecycle-wide responsibility should be viewed through dual perspectives: \textit{subjective} and \textit{objective}, which are complementary dimensions essential to ensure that system behaviors align with well-established responsible criteria in MAS, such as SMART (Specific, Measurable, Achievable, Relevant, and Time-bound) \cite{doran1981there, bjerke2017being}. The subjective perspective encompasses \textit{cognitive and intentional dimensions} (e.g., ethical awareness, accountability, reflexivity), which demand \textit{human-oriented approach} to reflect value-laden reasoning. Simultaneously, the objective perspective is rooted in \textit{quantifiable and verifiable elements} (e.g., uncertainty quantification, formal guarantees), requiring solid theoretical foundations and rigorous definitions \cite{huang2023survey}. These perspectives are intrinsically interdependent—subjective aspects of mutual understanding between agents require objective metrics to offer provable guarantees, while objective metrics need subjective modeling to provide guidance consistent with human values and requirements. For instance, LLM-based agents may exhibit emergent behaviors, requiring objective \textit{trust scores} to support subjective \textit{trustworthiness}. Meanwhile, cascading uncertainties across agent networks demand both \textit{technical solutions} and \textit{human judgment} to determine when system integrity is compromised.
Our core position is that \textbf{\textit{building responsible LLM-MAS requires a paradigm shift: from local, superficial agent-level alignment to global, systemic agreement.}} This shift rests on the following constitutive principles:
\begin{enumerate}
   \item \textbf{Agents must understand one another} through subjective interpretation while relying on objective metrics with guarantees to assess global inter-agent agreement.
   

   \item \textbf{Objective technical mechanisms (e.g., uncertainty quantification) are essential}, but must be guided by human values and societal expectations to define ethically aligned and task-specific requirements.
   
   \item \textbf{Systemic responsibility requires human and artificial intelligence (AI) co-moderation:} Human-centered moderators address ethical dilemmas and exceptional cases, while AI moderators uphold responsible behavior during routine operations.
\end{enumerate}
This paper reviews existing techniques for agreement (Section 2), uncertainty (Section 3), and adversarial vulnerabilities (Section 4) in LLM-MAS, identifying key limitations and analyzing their implications from dual perspectives. We then advocate for a responsible LLM-MAS design incorporating interdisciplinary principles and meta-governance (Section 5). Finally, we discuss alternative viewpoints (Section 6) and summarize our position (Section 7).

\begin{table*}
\caption{Responsibility in practical LLM-MAS applications. Symbols denote levels of support: 
\cmark~indicates support, \dash~indicates partial or limited support, and \xmark~indicates lack of support.}
\label{tab:llmmas_responsibility}
\scriptsize
\centering
\begin{tabularx}{\textwidth}{
  >{\raggedright\arraybackslash}p{0.21\textwidth}   
  >{\centering\arraybackslash}p{0.06\textwidth}    
  >{\centering\arraybackslash}p{0.07\textwidth}    
  >{\centering\arraybackslash}p{0.04\textwidth}    
  >{\centering\arraybackslash}p{0.07\textwidth}    
  >{\raggedright\arraybackslash}X                  
}
\toprule
\textbf{Application} & \textbf{Agreement} & \textbf{Uncertainty} & \textbf{Security} & \textbf{Governance} & \textbf{Responsibility Gap Explanation} \\
\midrule
Software Development \cite{qian-etal-2024-chatdev} & \dash & \xmark & \xmark & \dash & Prompt-based coordination with limited human oversight; no uncertainty and safeguards \\
Industrial Automation \cite{xia2024control} & \dash & \xmark & \xmark & \dash & Prompt-based coordination with limited AI oversight; no uncertainty and safeguards \\
Embodied Agents \cite{fan2025embodied} & \dash & \xmark & \xmark & \xmark & Prompt-based coordination; no uncertainty, safeguards and oversight consideration \\
Scientific Discovery \cite{ghafarollahi2024sciagents} & \dash & \dash & \xmark & \dash & Prompt-based coordination; post-hoc uncertainty scoring; no safeguards; limited human-AI oversight \\
Debate \cite{liang2023encouraging} & \dash & \dash & \xmark & \dash &  Prompt-based coordination with limited uncertainty measure and AI oversight; no safeguards \\
Gaming \cite{li2024llm} & \dash & \dash & \xmark & \xmark & Prompt-based coordination with limited uncertainty measure; no safeguards and oversight \\
Societal Simulation \cite{piao2025agentsociety} & \cmark & \xmark & \xmark & \cmark & Prompt-based, interdisciplinary-driven agents with mind-behavior modeling and human-AI oversight; no uncertainty and safeguards consideration \\
Financial Trading \cite{lin2024strategic} & \dash & \xmark & \xmark & \xmark & Prompt-based coordination; no uncertainty, safeguards and oversight consideration \\
Recommend Systems \cite{fang2024multi} & \dash & \xmark & \xmark & \xmark & Prompt-based coordination; no uncertainty, safeguards and oversight consideration \\
Disease Simulation \cite{williams2023epidemic} & \dash & \xmark & \xmark & \dash & Prompt-based coordination with post-hoc logging oversight; no uncertainty and safeguards  \\
\bottomrule
\end{tabularx}
\vspace{-15pt}
\end{table*}
\section{Agreement in LLM-MAS}
In LLM applications, research typically employs the concept of \textit{alignment}, which ensures an LLM’s conformity to objectives such as ethical norms, human intentions, or task-specific requirements and has proven reasonable for single-agent scenarios. Nevertheless, in LLM-MAS, where multiple heterogeneous agents interact dynamically, alignment alone is insufficient to capture the system-level behaviors and relationship required for responsible MAS. Hence, we reconceptualize \textbf{agreement} as an extension of alignment that incorporates its objectives while places particular emphasis on \textit{providing system-level guarantees of behavioral coherence and inter-agent mutual understanding, including coordinated outputs, trustworthy collective decisions, and unified semantic interpretations.}

\subsection{Agreement Formation}
The reconceptualization of agreement in LLM-MAS requires us rethinking system-wide responsibility beyond traditional protocol-based designs. For achieving global agreement among heterogeneous agents, the responsible design requires jointly addressing both subjective mutual understanding and objective probabilistic guarantees \cite{xu2023reasoninglargelanguagemodels, zhao-etal-2024-electoral}. Although recent methods align agents with human preferences and improve inter-agent coordination, most studies \cite{kirchner2022researchingalignmentresearchunsupervised, shen2023largelanguagemodelalignment, cao2024scalableautomatedalignmentllms, pan2023automaticallycorrectinglargelanguage, fernandes-etal-2023-bridging} focus on local, single-agent alignment rather than ensure global agreement in complex LLM-MAS. To better analyze these design, we review them into agent-human and agent-agent in the following sections.
\subsubsection{Agent to Human Agreement}
To establish agreement with human values or preferences, LLM agents must accurately interpret natural language, 
grasp assigned requirements or goals, 
and understand societal regulations. 
The existing solutions can be broadly classified
into reinforcement learning, supervised fine-tuning, and self-improvement with graphical illustrations shown in Appendix~\ref{app:agent_to_human}.

\textbf{Reinforcement Learning}
is the most commonly used method to achieve human preference agreement via human feedback (RLHF) \cite{ouyang2022training, stiennon2020learning, ziegler2019fine}.
RLHF typically fine-tune models' policies using reward models derived from human or LLM-generated feedback (e.g., PPO algorithm \cite{schulman2017proximalpolicyoptimizationalgorithms} or automated preference labels \cite{bai2022constitutionalaiharmlessnessai, lee2024rlaifvsrlhfscaling}), with continued improvements in \cite{glaese2022improvingalignmentdialogueagents, bai2022traininghelpfulharmlessassistant, tan-etal-2023-self, kirk-etal-2023-past, zhu2024principledreinforcementlearninghuman}.


\textbf{Supervised Fine-tuning (SFT) }
\cite{dong2023raftrewardrankedfinetuning, alpaca2023} 
aligns agents with human preferences by minimizing the loss between model outputs and labeled data, such as instruction-response pairs \cite{taori2023alpaca, ding-etal-2023-enhancing} and query-based preferences \cite{guo2024controllablepreferenceoptimizationcontrollable}. Among them, Instruction-finetuning (IFT) normally targets static tasks, while preference labeling captures dynamic user needs. 
Models like Stanford Alpaca \cite{alpaca2023}, AlpaGasus \cite{chen2024alpagasustrainingbetteralpaca}, and InstructGPT \cite{ouyang2022training} demonstrate improved alignment through SFT. Recent frameworks, including LIMA \cite{zhou2023limaalignment}, PRELUDE \cite{gao2024aligningllmagentslearning}, and the Preference Tree \cite{yuan2024advancingllmreasoninggeneralists}, further refine this process by learning user preferences from prompts, dialogues, and reasoning paths. \\
\vspace{-10pt}
\textbf{Self-improvement}
refines agreement by leveraging inductive biases. 
It applies self-consistency \cite{wang2023selfconsistencyimproveschainthought} through Chain-of-Thought (CoT) \cite{wei2022chain} or Tree-of-Thought (ToT) \cite{yao2024tree}, selecting the most consistent reasoning paths to improve output quality. Based on this, Self-Improve \cite{huang2022largelanguagemodelsselfimprove} fine-tunes models using high-confidence inferences. SAIL \cite{ding2024sailselfimprovingefficientonline} combines SFT and 
online RLHF to reduce dependence on human feedback by bi-level optimization. Moreover, self-rewarding \cite{yuan2024selfrewardinglanguagemodels} and Meta-Judge \cite{wu2024metarewardinglanguagemodelsselfimproving} assess and enhance model outputs by autonomously optimizing judgment skills.

\subsubsection{Agent to Agent Agreement}
In MAS, agreement arises when agents consistently interpret and respond to each other's intents, information, and outputs, enabling coherent collective decisions \cite{10720863}. This section reviews existing agreement approaches among heterogeneous agents, with graphical examples in Appendix~\ref{app:agent_to_agent}.

\textbf{Cross-Model Agreement}
follows two main directions.
Strong-to-weak approaches use an aligned teacher model to generate response pairs \cite{xu2024wizardlm, taori2023alpaca, peng2023instructiontuninggpt4} and preferences \cite{cui2024ultrafeedbackboostinglanguagemodels} for training unaligned model. For example, Zephyr utilizes distilled  SFT \cite{tunstall2023zephyrdirectdistillationlm} with teacher models acting as automated labeler before final preference optimization via DPO.
Weak-to-strong method, like SAMI \cite{fränken2024selfsupervisedalignmentmutualinformation} and others in \cite{burns2023weaktostronggeneralizationelicitingstrong}, use weak models to generate guiding signals for stronger ones.
Additionally, information-theoretic techniques, such as PMIC \cite{li2023pmicimprovingmultiagentreinforcement}, promote alignment by maximizing mutual information among superior behaviors and minimizing it among inferior ones.

\textbf{Task Decomposition} aims to break complex tasks into relatively easy sub-tasks to form agreement. Based on this intuition, Factored Cognition \cite{10.5555/3495724.3495977} decomposes tasks into numerous smaller, independent components for separate processing. IDA \cite{christiano2018supervisingstronglearnersamplifying} similarly splits large tasks into simpler ones: during amplification, human experts decide how to answer a question after reviewing the model’s sub-answers; in distillation, the model learns how to both solve and decompose tasks. This process resembles AlphaGo Zero \cite{addanki2019understanding}, where recursive self-play refines both strategy and task breakdown. In LLM-MAS, the roles of human experts and subtask solvers can be alternated among LLM agents, enabling a collective understanding of the overall task. MetaGPT \cite{hong2024metagptmetaprogrammingmultiagent} and AgentVerse \cite{chen2023agentversefacilitatingmultiagentcollaboration} adopt this paradigm by assigning subtasks to multiple agents, forming an assembly-line paradigm.


\textbf{Debate and Adversarial Self-Play}
refine agreement by leveraging adversarial dynamics, 
especially in interdisciplinary MAS. There normally adopt two frameworks: Generator-Discriminator and Debate. In the Generator-Discriminator framework, the generator generates the
response, and the discriminator judges the quality. For instance, CONSENSUS GAME \cite{jacob2023consensusgamelanguagemodel} iteratively adjusts both agents’ policies to minimize regret and achieve a regularized Nash equilibrium. Regarding the Debate, models engage in argumentation to strengthen 
reasoning and alignment. For example, in \cite{irving2018aisafetydebate}, supervised pre-trained models act as debaters, generating arguments that withstand critique, with RLHF guiding the debate toward a Nash equilibrium aligned with human expectations.


\textbf{Shared Aligners} \cite{ngweta2024alignersdecouplingllmsalignment} separate alignment from LLMs via modular aligners and inspectors. The LLM produces initial outputs, aligners optimize them based on task-specific objectives, and inspectors selectively trigger aligners based on evaluation to saves human annotation costs.


\textbf{Environment Feedback} fosters interdisciplinary agreement by integrating multimodal background knowledge into a World Model (WM) \cite{lecun2022path}, enabling agents to handle diverse tasks and roles under shared common sense. Agents’ states and actions serve as inputs, while the WM predicts possible state transitions, probabilities, and relative rewards, guiding long-term optimal strategies \cite{hu2023languagemodelsagentmodels}. Environment-driven tasks may also invoke external tools or social simulators to extend agreement
to task-specific scenarios. For example, MoralDial \cite{sun2023moraldialframeworktrainevaluate} simulates social discussions between agents and the environment, enhancing agents’ moral reasoning, explanation, and revision. 

\textbf{Barriers to Agreement in LLM-MAS:} 
Despite recent advances in agreement-formation techniques within LLM-MAS, agents inevitably encounter \textbf{conflicts} from both subjective divergences (e.g., interpretations, ethical values) and objective asymmetries (e.g., knowledge, goals) \cite{phelps2023models}. For instance, the inherent semantic ambiguities of natural language amplify misalignments in interpretation. In autonomous driving scenarios, an alert to "slow down due to road conditions" may be understood differently by various agents, resulting in inconsistent implementations \cite{yang2024llmdrive}. Similarly, in collaborative planning, agents may prioritize competing objectives like performance versus safety, leading to divergent trade-offs strategies \cite{tessier2005conflicting}. Knowledge-based conflicts in LLM-MAS further arise from variations in reasoning paths and information sources, where agents may construct distinct mental models or reach contradictory conclusions, even when starting from identical initial information \cite{wang2024astute,DBLP:journals/corr/abs-2407-07791}. Such conflicts also result in \textbf{hallucination}, defined as the generation of fluent yet factually incorrect information. Hallucination introduces significant systemic risks in MAS due to the probabilistic nature of LLMs and conflicts between LLM agents \cite{ji2023survey}. When hallucinated information from one agent is accepted as valid by others, it creates propagation cycles where misinformation is reinforced through repeated interactions and is hard to be detected \cite{huang2023survey}. This vulnerability is particularly concerning as adversaries could exploit it to transform local disagreements into system-wide failures. Thus, while LLM-based agents offer advantages, they also pose a new dilemma, which is, \textit{we must determine whether integrating LLMs into MAS can prevent conflicts from inherent knowledge ambiguities in LLM and produce outcomes aligned with our expectations.}

\textbf{Toward Responsible Agreement via Dual-Perspective Integration:}
Current approaches rely mainly on ad-hoc solutions \cite{bhatia2020preference,liu2024autonomous,din2024ontology}, which lack robust mechanisms to quantify and validate disagreement in LLM-MAS. As a result, conflicts may be obscured when agents operate with imperfect agreement, often leading to overconfident yet unreliable decisions \cite{rodriguez2023good}.
In contrast, we advocate a principled framework extending the Belief-Desire-Intention (BDI) framework to integrate possible solutions from dual perspectives. The concept of BDI is originally introduced by Bratman in 1987 to model human reasoning and later formalized for artificial agents by Rao and Georgeff in the 1990s. It organizes decision-making into beliefs (environmental information), desires (goals), and intentions (action commitments) \cite{bratman1987intention,rao1997modeling}.
Our potential extension incorporates hierarchical conflict resolution. The belief layer applies formal verification to standardize the interpretation of ambiguous instructions. The knowledge layer, aligned with desires, uses probabilistic belief updating, such as Conformal Bayesian Inference \cite{fong2021conformal}, to weigh conflicting information by source reliability and context. The intention layer could employ uncertainty-aware multi-criteria decision theory to balance goals and ethical constraints in complex scenarios.
This framework can be further augmented with causal reasoning, applying causal models to proactively identify potential conflicts rather than responding reactively \cite{zeng2022multi}. In addition, guardrail techniques combined with RAG can be integrated to mitigate hallucinations and malicious manipulations, as demonstrated in real-world applications such as Amazon's virtual assistants \cite{dong2024buildingguardrailslargelanguage,hu2025trustorientedadaptiveguardrailslarge,dong2024safeguarding}. \textit{We view conflicts not as anomalies to be eliminated, but as inherent system features requiring dedicated management mechanisms with theoretical foundations.}

\vspace{-5pt}
\subsection{Agreement Evaluation}
\vspace{-5pt}
To determine whether an LLM-MAS is responsible, both subjective and objective evaluation methods are essential for assessing whether the system’s agreements meet acceptable standards. Many existing metrics and frameworks already incorporate both perspectives. For example, MAgIC \cite{xu2024magicinvestigationlargelanguage} introduces objective measures for cooperation and coordination, evaluating the success rate of achieving common goals. Opinion consistency and convergence \cite{li2023quantifyingimpactlargelanguage}, though based on measurable opinion shifts, also reflect subjective alignment among agents.
Trust scores \cite{fung2024trustbasedconsensusmultiagentreinforcement, decerqueira2024trustaiagentsexperimental} capture subjective trustworthiness that influence consensus formation. Consensus variance \cite{chen2025multiagentconsensusseekinglarge} provides an objective metric by quantifying differences in agents’ final states after negotiation. Finally, semantic similarity \cite{xu2024reasoningcomparisonllmenhancedsemantic, aynetdinov2024semscoreautomatedevaluationinstructiontuned} typically serves as an objective measure of agreement during optimization.

\textbf{The Evaluation Gap under Dynamic Multi-Agent Interactions:}
While single-agent evaluations and static task assessments rely on established metrics, evaluating agreement in LLM-MAS is substantially more complex. Multi-agent interactions involve evolving agreement states shaped by feedback loops and historical dependencies \cite{shen2023large}, leading to asymmetric belief updates \cite{schubert2024context} and divergent outcomes depending on interaction sequences \cite{yoffe2024debunc}. 
For example, in a planning scenario, if Agent A proposes a solution before Agent B, the resulting strategy may differ from the case where B proposes first, even with identical starting conditions. Moreover, the lack of unified frameworks for aggregating agent-level metrics \cite{guo2024large} further complicates system-wide assessment. Standard ethical metrics like toxicity filtering and bias detection capture individual agent performance but overlook irresponsible behaviors where local optimality results in global suboptimality. Even general metrics like response efficiency and task completion rates fail to capture the dual-perspective trade-offs between subjective ethical values and objective performance, especially as LLMs can exhibit selfish strategies like maintain dominant position during interaction \cite{tennant2024moral}, making optimality across all evaluation criteria inherently unattainable. Notably, \cite{wang2024rethinking} demonstrate that adding more capable agents can degrade collective performance, \textit{indicating that the participation of more individually well-performing agents does not necessarily lead to better outcomes.}

\textbf{Learning-Based Evaluation Method is Needed:} Current evaluation methods for agreement in LLM-MAS mainly emphasize \textit{static metrics extended from single agents}, often neglecting the \textit{dynamic evolution of multi-agent agreement during task execution}. Recent approaches try to use LLMs as dynamic evaluators also \textit{lack theoretical guarantees} and remain \textit{sensitive to subjective factors} like the prompt template design \cite{wei2024systematic}. We advocate a learning-based strategy that adapts to evolving agent interactions. Leveraging optimization techniques like \textit{metric learning} or \textit{submodular optimization}, which both aim to identify meaningful representations and efficiently select diverse, representative evaluation criteria \cite{huisman2021survey, chen2024less}, the system can synthesize subjective and objective evaluation functions (e.g., ethical alignment and task performance) to enable dynamic optimization of multi-dimensional agreement metrics based on real interaction patterns. By learning context-aware subspace projections, the framework supports \textit{probabilistic interpretability} of system responsibility \cite{liao2023reimagining}, offering transparent insights into the level of systemic agreement.
\vspace{-5pt}
\section{Uncertainty in LLM-MAS} 
As LLM-MAS shifts from single-agent planning to multi-agent collaboration, uncertainty management is vital for both technical reliability and human-aligned responsibility. While traceability and probabilistic guarantees are essential, quantitative measures alone are insufficient without grounding in human values that define acceptable risks and ethical boundaries. This section examines internal and external uncertainties in LLM-MAS and emphasizes the need for probabilistic, traceable, and human-aligned uncertainty management to maintain continuous global agreement and responsibility.

\subsection{Uncertainty in LLM Agents' Internal Mechanism}
In many widely deployed LLM applications, even a single agent can often be viewed as a complex \textit{multi-agent or multi-component system.} Decomposing such LLM agents into functional components often involves additional uncertainties in memory management and strategic planning module beyond traditional agent, both of which must be designed to uphold responsibility.

\textbf{Memory} 
RAG enhances LLM agents' memory ability by integrating external, up-to-date, domain-specific knowledge, improving factual accuracy and reducing hallucinations without extensive retraining. However, not all retrieved sources equally influence decision-making. To address this, attention-based uncertainty quantification ~\cite{duan2024shifting} estimates uncertainty via variance in attention weights across retrieved sources, while LUQ~\cite{zhang2024luq} uses an ensemble approach to re-rank documents and adjust verbosity based on confidence. Xu et al.~\cite{xu2024unsupervised} further introduce a self-consistency mechanismthat compares retrieved evidence with generated outputs to refine both retrieval and generation.

\textbf{Planning} Planning enables structured decision-making by decomposing complex tasks into manageable steps but remains a major source of uncertainty, especially in stochastic environments. Recent methods mitigate this by enhancing both prediction accuracy and decision confidence. Tsai et al.\cite{tsai2024efficient} fine-tuned Mistral-7B to evaluate prompt-action compatibility with conformal prediction. Ren et al.\cite{ren2023robots} introduced KnowNo, estimating token probabilities to flag when human oversight is necessary. IntroPlan~\cite{liang2024introspective} builds upon these by implementing introspective planning with tighter confidence bounds, reducing reliance on human intervention.

\subsection{Uncertainty in LLM Agents' External Interaction}
While uncertainty quantification in LLM agents has received growing attention, existing methods mostly operate at the instance level, neglecting multi-turn interaction history. This limitation is especially critical in real-world settings like autonomous medical assistants \cite{li2024mediq, savage2024large}, where agent responses must consistently reflect evolving contextual dependencies \cite{chen2024llmarena, pan2024agentcoord}. In MAS setting, recent methods have begun integrating uncertainty management into agent networks. DiverseAgentEntropy \cite{feng2024diverseagententropy} quantifies uncertainty by probing factual knowledge under black-box constraints, improving prediction accuracy and exposing retrieval failures. In contrast, DebUnc \cite{yoffe2024debunc} incorporates confidence metrics into multi-turn reasoning via attention adjustment and prompt-based signaling. Although such approaches embed uncertainty into decision-making, their lack of theoretical guarantees and disregard for the appropriateness of uncertainty measures may introduce side effects, that is, \textit{irresponsible uncertainty measures can actually undermine agreement across heterogeneous agents.}

\subsection{Uncertainty across the LLM-MAS}
While a variety of statistical techniques have been developed to estimate uncertainty in LLMs, most remain focused on \textbf{single models or isolated outputs and have yet to scale to holistic uncertainty management across the entire LLM-MAS.} Existing methods primarily rely on statistical analysis, including single-inference approaches based on token log probabilities~\cite{yang2023improving,ren2023robots}, and multi-inference methods that assess semantic consistency across multiple outputs based on semantic equivalence learned through unified concepts, such as semantic entropy~\cite{farquhar2024detecting} and spectral clustering~\cite{lin2023generating}. Beyond purely statistical methods, ``Human-in-the-loop'' has been advocated by the MAS community, which is preliminarily introduced into uncertainty management. For example, KnowLoop~\cite{zheng2024evaluating} uses entropy-based measures to flag unpredictable outputs for human review, while UALA~\cite{han2024towards} applies maximum or mean uncertainty metrics to identify knowledge gaps and trigger clarification.

\textbf{Limitations on Applicability:}
Although this progress partially supports our position that objective quantification must be guided by human values and has shown improved performance, \textit{LLM-MAS still lacks rigorous quantitative methods that incorporate traceable agent interaction histories to monitor runtime uncertainty and establish verifiable statistical bounds for system-wide responsibility.} These limitations constrain the applicability of existing methods to real-world scenarios~\cite{yang2024confidence,DBLP:journals/corr/abs-2407-11282}.

\textbf{The Ripple Effects of Imperfect Uncertainty Management:} This aforementioned gap amplifies two unique challenges in LLM-MAS: \textit{"knowledge drift"}, which refers to the gradual deviation of agent reasoning from factual accuracy during multi-turn interactions and \textit{"misinformed perspective propagation"}, where incorrect information spreads throughout the system and overrides others' viewpoints. Unlike traditional MAS with explicitly programmed rules, these issues emerge from the inherent variability of natural language processing~\cite{fastowski2024understanding,xu-etal-2024-earth,wang2024rethinking} and intensify during collaborative reasoning. In such settings, agents tend to align with incorrect consensus (conformity effect) or defer to perceived authority, distorting knowledge bases even when some agents initially hold correct ideas \cite{zhang2024exploring}. For example, in multi-agent debates, an agent with flawed understanding may create persuasive yet erroneous rationales that divert the entire group away from accurate solutions \cite{breum2024persuasive}. Additionally, LLM agents exhibit a tendency for expanding their bias, wherein, unlike humans who filter mistakes, they amplify errors, further exacerbating knowledge drift and collective reasoning inaccuracies \cite{liu2024exploring}. Existing approaches, such as prompt engineering \cite{fernando2024promptbreeder}, the use of LLM agents as judge to arbitrate and refine reasoning \cite{zheng2023judging,chan2024chateval}, and "human-in-the-loop" intervention \cite{triem2024tipping}, attempt to address these issues. However, prompt engineering often lacks scalability and struggles with context-specific requirements, while purely human intervention is labour-intensive and impractical for large-scale systems. Moreover, judge agents, being LLM-based themselves, are susceptible to similar biases and can unintentionally reinforce reasoning errors, leaving a persistent issue \cite{wang2024rethinking}.

\textbf{Toward Provable System-Wide Uncertainty Management:} Addressing aforementioned challenges requires moving beyond heuristic solutions toward principled system architectures with \textit{provable guarantees} that integrate both \textit{objective} probabilistic assurances and \textit{subjective} value-guided modeling~\cite{bensalem2023indeed}. We advocate a probabilistic-centric architecture that not only quantifies and propagates uncertainty across the agent network but also incorporates comprehensive human-aligned modeling to assess whether uncertainties across agents or components meet subjective requirements and societal values, such as acceptable risk thresholds. Specifically, LLM-MAS should: (1) implement rigorous probabilistic frameworks for \textbf{\textit{uncertainty propagation}} to maintain consistency; (2) establish formal verification mechanisms providing \textbf{\textit{statistical}} or \textbf{\textit{deterministic bounds}} on knowledge drift; and (3) \textbf{\textit{Modeling LLM-MAS state transitions}} to formally monitor system-wide runtime uncertainty and agreement with human objectives and contextual constraints, similar as automata theory and ontology engineering \cite{kallwies2022symbolic,giannoulis2018cosmos}. For example, although conformal prediction~\cite{wang2024probabilistically} has yet to be widely applied to system-wide multi-agent systems in safety-critical domains, it exemplifies the integration of probabilistic guarantees that align collective decisions with specified confidence levels while accounting for individual agent uncertainties, guided by human-centered acceptability criteria.
\section{Malicious Exploits and Safety Threats in LLM-MAS} 
This section examines critical security vulnerabilities in LLM-MAS, highlighting adversarial threats that can jeopardize agent-level safety and compromise systemic agreement. We further outline our perspectives and potential solutions to address these risks.

\subsection{Collusive Behavior}
Collusion in LLM-MAS refers to the coordinated behavior among two or more agents that serves their mutual benefit while potentially undermining system objectives, fairness, or established constraints \cite{huang2024survey}. This includes: (1) \textit{explicit collusion} through direct exchange of strategic information, and (2) \textit{implicit collusion} where agents independently converge on cooperative strategies without direct communication. 
For instance, research has demonstrated that LLM agents in Cournot competition can engage in implicit collusion, such as covert market division without explicit coordination or instructions, thereby evading detection \cite{wu2024shall,lin2024strategic}.
Furthermore, semantic cues or steganographic techniques further support collusive behaviors, \textit{making them hard to identify and easily exploitable by adversaries} \cite{motwani2024secret}.
LLM's opaqueness further exacerbates the issue, as their outputs are often contextually plausible, effectively masking the underlying collusive dynamics.
\subsection{Data Poisoning \& Jailbreaking Attack}
Data poisoning and jailbreaking attacks introduce additional vulnerabilities in LLM-MAS by exploiting communication channels, contaminated knowledge retrieval, and manipulated context windows \cite{das2024security}. Unlike conventional industrial application, where poisoning only targets the training phase and can often be mitigated through data sanitization \cite{koh2022stronger} or adversarial testing \cite{11077439}, LLM-MAS faces expanded attack vectors due to its reliance on dynamic interactions and external knowledge \cite{das2024security}. For instance, 
RAG may inadvertently allow poisoned knowledge bases to infiltrate the system \cite{chen2024agentpoison}. Furthermore, natural language communication between agents further amplifies the attack surface, allowing adversaries to exploit LLMs' context sensitivity through subtle linguistic manipulations and safety-bypassing prompts. Jailbreaking, originally aimed at bypassing safety constraints in individual LLMs and now extending across modalities, becomes more dangerous in LLM-MAS \cite{liu2024jailjudge, yin2025taijitextualanchoringimmunizing}. The property of misinformation propagation leads to both poisoned and jailbroken information being enhanced through collaborative reasoning, creating cascading security breaches across the system. 


\subsection{Cyber Threats}
Cyber threats 
also become a significant challenge to LLM-MAS due to their distributed architecture and complex interaction patterns \cite{zeeshan2025large}. Network-level attacks, such as wormhole \cite{ren2024hwmp} and denial-of-service \cite{wen2023secure}, can disrupt temporal consistency and degrade operational performance. The frequent API interactions required for LLM services and inter-agent communication not only expose vulnerabilities in network protocols and authentication mechanisms, but also create performance bottlenecks \cite{wang2024large}. Furthermore, the integration of external knowledge sources introduces more attack targets \cite{gummadi2024enhancing}, highlighting the need for robust cybersecurity measures that \textit{balance protection with system responsiveness, while quantifying the timeliness and completeness of information exchange.}

\textbf{The Inadequacy of Static Protections in LLM-MAS:} Current mitigation strategies, while effective for individual LLM, face limitations when extended to LLM-MAS. 
For collusive behaviors, existing detection mechanisms rely heavily on post-hoc log analysis, which fails to support real-time intervention in dynamic LLM-MAS  \cite{bonjour2022information,motwani2024secret}. Similarly, data poisoning and jailbreaking defenses primarily focus on robust training and input sanitization at model initialization, becoming inadequate as compromised information can be injected and propagate through various interaction channels during runtime \cite{wang2022threats}. Traditional cybersecurity measures like rule-based firewalls, struggle to address both the uncertainties from dynamic reasoning and the increased communication channels in LLM-MAS \cite{applebaum2016firewall}. Moreover, network-level detection have proven less effective against LLM-generated misinformation, which often appears more deceptive despite being semantically similar to human-crafted attacks \cite{chen2024can}. These  \textit{static protection} are insufficient for dynamically protecting knowledge exchange and accumulation in interactive MAS,
\textit{underscoring the necessity for 
dedicated runtime mechanisms to continuously detect and filter potentially compromised data while preserving global agreement.}

\textbf{Toward Responsible Runtime Oversight under Adversarial Threats:}
We suggest a runtime monitoring and AI provenance framework enhanced by uncertainty-based governance rules \cite{souza2022workflow,werder2022establishing,xu2022dependency}. This approach enables continuous surveillance of system behaviors, tracking information flow and decision origins. By integrating provenance chain modeling, uncertainty quantification, or certified robustness techniques such as randomized smoothing, the system can dynamically trace and validate information propagation with probabilistic guarantees \cite{shorinwa2024survey, HU2025103779}. Besides, the framework should enable adaptive monitoring that dynamically adjusts scrutiny based on risk or reputation, maintaining reliable and interpretable records of decisions \cite{hu2025trustorientedadaptiveguardrailslarge}. Also, machine unlearning can remediate contaminated representations \cite{pawelczyk2024incontext,hu2025falconfinegrainedactivationmanipulation}, while neural-symbolic methods combine symbolic reasoning (e.g., abductive inference) with neural flexibility to enhance safety and verifiability \cite{tsamoura2021neural}. Embedding these capabilities into the system architecture enables both security and transparency, ensuring robust and trustworthy operation under uncertainty.

\section{Interdisciplinary Integration and Meta-Governance for LLM-MAS}
\textbf{Interdisciplinary Guidance for Reducing Integration Hurdles:}
Achieving responsible LLM-MAS requires interdisciplinary guidance to reduce integration hurdles and systematically combine subjective modeling with objective verification across diverse technologies \cite{gao2024large}. By aligning domain knowledge, regulatory requirements, and technical principles, interdisciplinary insights help LLM-MAS more effectively address challenges in agreement formation, uncertainty management, and resilience under adversarial conditions \cite{handler2023balancing}. Crucially, this guidance underpins human-centered lifecycle design, informing how diverse human roles are structured and integrated to ensure consistent agreement with both technical and ethical objectives.
\begin{figure}[htbp]
    \vspace{-10pt}
    \centering
    \includegraphics[width=1\columnwidth]{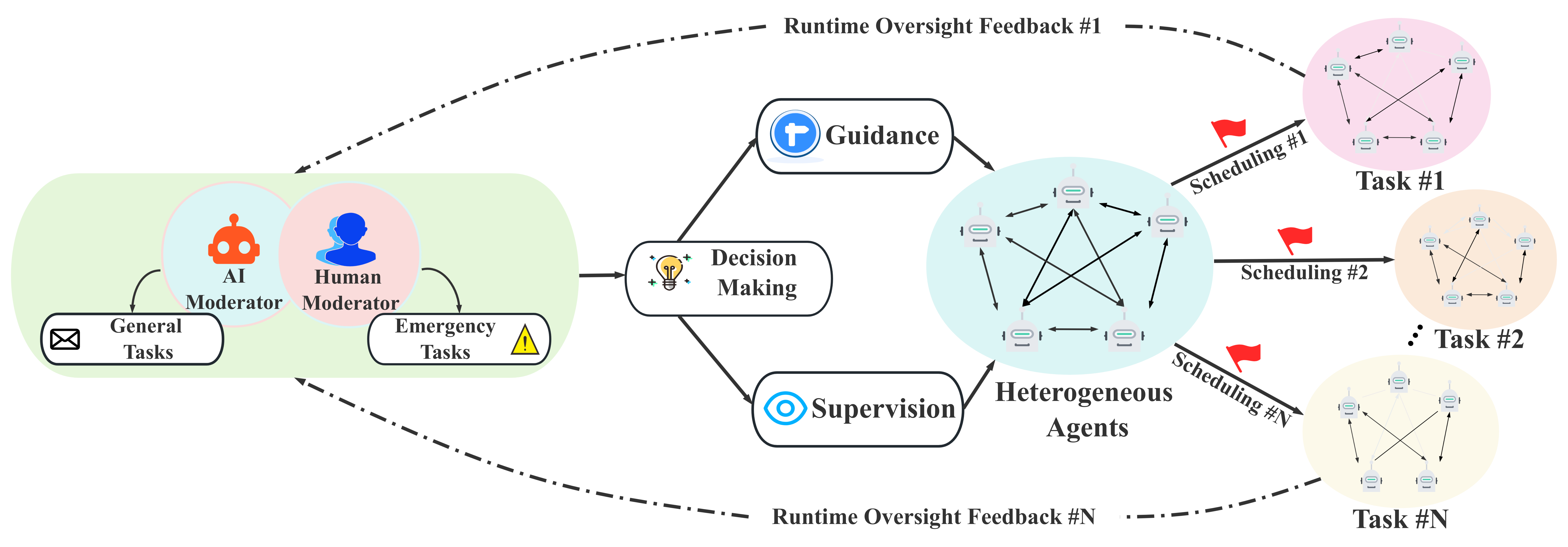}
    \caption{Illustration of Responsible LLM-MAS Framework}
    \label{safeframework}
    \vspace{-5pt}
\end{figure}

\textbf{Four-Stage Human-Centered Lifecycle Design:} While human-in-the-loop concepts are widely referenced across various fields, they rarely provide explicit role definitions for human involvement at each lifecycle stage. We address this gap by specifying structured human responsibilities that complement automated processes. Accordingly, we propose a human-centered lifecycle design defining human involvement across four critical stages: (1) In the design stage, human regulators translate ethical principles into technical specifications and establish acceptance boundaries. (2) During development, human experts implement uncertainty quantification, agreement verification, and safety guarantees. (3) At deployment, human evaluators assess real-world performance and validate agreement with user needs and societal standards. (4) During maintenance, human monitors oversee system behavior, identify emergent patterns, and adapt the system to evolving requirements.

\textbf{Human-AI Collaborative Meta-Governance:} Based on these defined roles, we advocate that AI and human moderators collaborate as a meta-governance layer as shown in Figure~\ref{safeframework}. This synergy combines AI’s efficiency in routine monitoring with human oversight to mitigate the risks of overreliance that could otherwise result in catastrophic failures \cite{nikolaidis2004comparison}. Human moderators retain the ability to detect emerging issues and enact adaptations to address evolving requirements and facilitate system recovery for systemic responsibility \cite{benner2021you}. Furthermore, this joint collaboration can continuously monitor MAS, manage integration of evolving technologies, and dynamically evaluate whether system behaviors and designs remain adequate under changing environments. 

\section{Alternative Views}
\textbf{Feasibility concern of applying formal methods to LLMs}
The application of formal verification to LLMs faces scalability challenges due to their large parameter space, stochastic behavior, and lack of interpretability \cite{katz2017reluplex, huang2024survey}. These difficulties are compounded in LLM-MAS, where agent interactions evolve over time, making full-system verification impractical. Recent advances suggest more tractable alternatives. A prototypical implementation is the doubly-efficient debate protocol \cite{pmlr-v235-brown-cohen24a}, which avoids verifying internal model logic and instead structures oversight as a game between diverse LLM agents, where a lightweight verifier adjudicates based on their natural-language arguments. The protocol ensures that honest agents can win using only polynomial-time strategies, even against dishonest opponents with unbounded computation. Crucially, the verifier only queries a constant number of critical steps, making the overall oversight process efficient, scalable, and compatible with human supervision at any time. By shifting verification to the level of reasoning traces and interactive proof, this approach enables formal guarantees without incurring prohibitive runtime costs, supporting real-time deployment in dynamic LLM-MAS scenarios.

\textbf{Uncertainty quantification can accurately capture semantic gap}
Uncertainty quantification in LLM-MAS, if grounded solely in internal probability estimates, may fail to capture or detach the contextual nuances essential for responsible behavior, especially in ambiguous or value-laden settings \cite{kadavath2022language}.  To address such gaps, Kuhn et al. \cite{kuhn2023semantic} argue that epistemic uncertainty in large language models exhibits properties that defy conventional quantification methods, and propose a more reliable semantic uncertainty framework that aligns more closely with human semantic judgments. In contrast, Truthful AI \cite{evans2021truthful} shifts the focus from internal confidence to external accountability, advocating for standards based on the avoidance of negligent falsehoods rather than precise probabilistic calibration. Together, these perspectives underscore the need for subjective uncertainty modeling that is not only statistically coherent but also semantically meaningful and socially grounded.

\section{Conclusion} 


This position paper advocates that responsible LLM-MAS must move beyond isolated alignment techniques toward lifecycle-wide, system-level agreement grounded in dual perspectives. By integrating subjective human-centered values with objective verifiability, and embedding collaborative human-AI oversight throughout the system’s design, development, deployment, and maintenance stages, we can better manage uncertainty, ensure ethical coherence, and mitigate latent security hazards. This dual-perspective design enables scalable, auditable, and verifiable oversight in dynamic multi-agent interactions. Ultimately, we advocate reconceptualizing LLM-MAS as unified socio-technical systems requiring principled responsibility mechanisms to maintain trustworthy, resilient behavior in open and uncertain environments.

\bibliographystyle{plain}
\bibliography{Styles/reference}






\newpage
\appendix
\section{Existing Agreement Solutions between Agent and Human} \label{app:agent_to_human}

\begin{figure}[htbp]
    \centering
    \begin{subfigure}[t]{.9\textwidth}
        \centering
        \includegraphics[width=\textwidth]{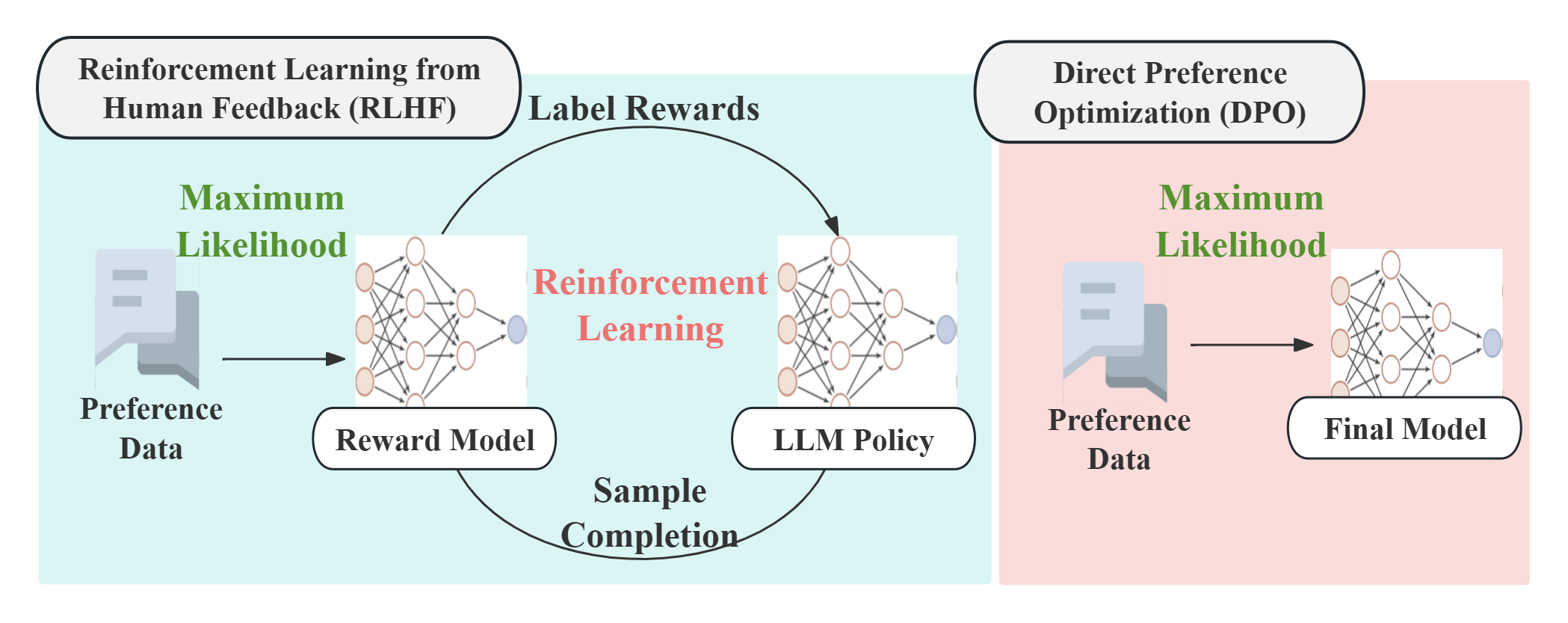}
        \caption{Reinforcement Learning Framework}
        \label{fig:RL}
    \end{subfigure}
    
    \vspace{8pt} 

    \begin{subfigure}[t]{.9\textwidth}
        \centering
        \includegraphics[width=\textwidth]{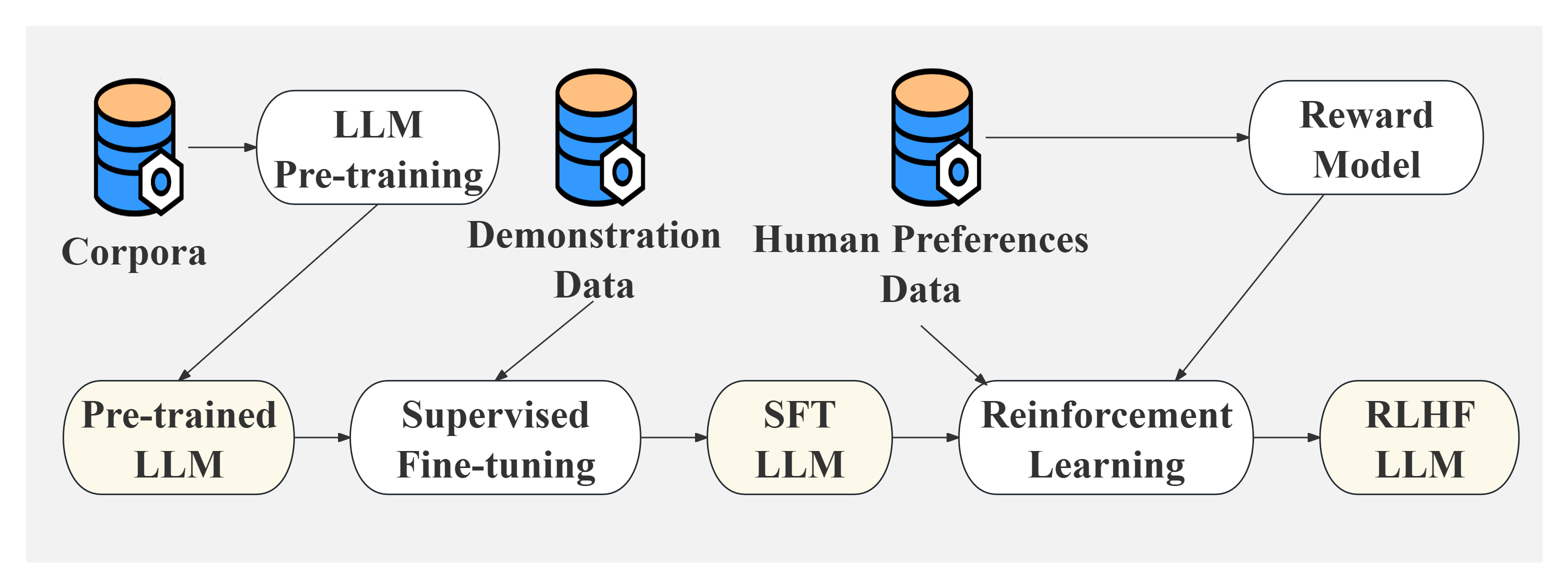}
        \caption{Supervised Fine-tuning}
        \label{fig:SFT}
    \end{subfigure}
    
    \vspace{8pt}

    \begin{subfigure}[t]{.9\textwidth}
        \centering
        \includegraphics[width=\textwidth]{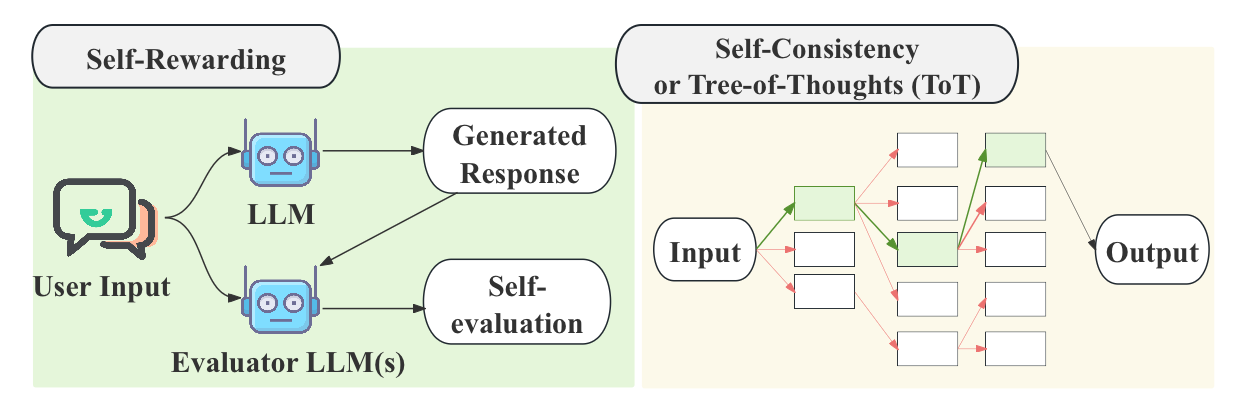}
        \caption{Self-Improvement Examples}
        \label{fig:SelfImprovement}
    \end{subfigure}
    \vspace{15pt}
    \caption{Taxonomy and Graphical Illustration of Agent-to-Human Agreement Techniques.}
    \label{fig:LearningParadigms}
\end{figure}




\newpage
\section{Existing Agreement Solutions between Agent and Agent} \label{app:agent_to_agent}
\begin{figure}[htbp]
    \vspace{-20pt}
    \centering
    \begin{adjustbox}{center}
        \includegraphics[width=1.05\textwidth]{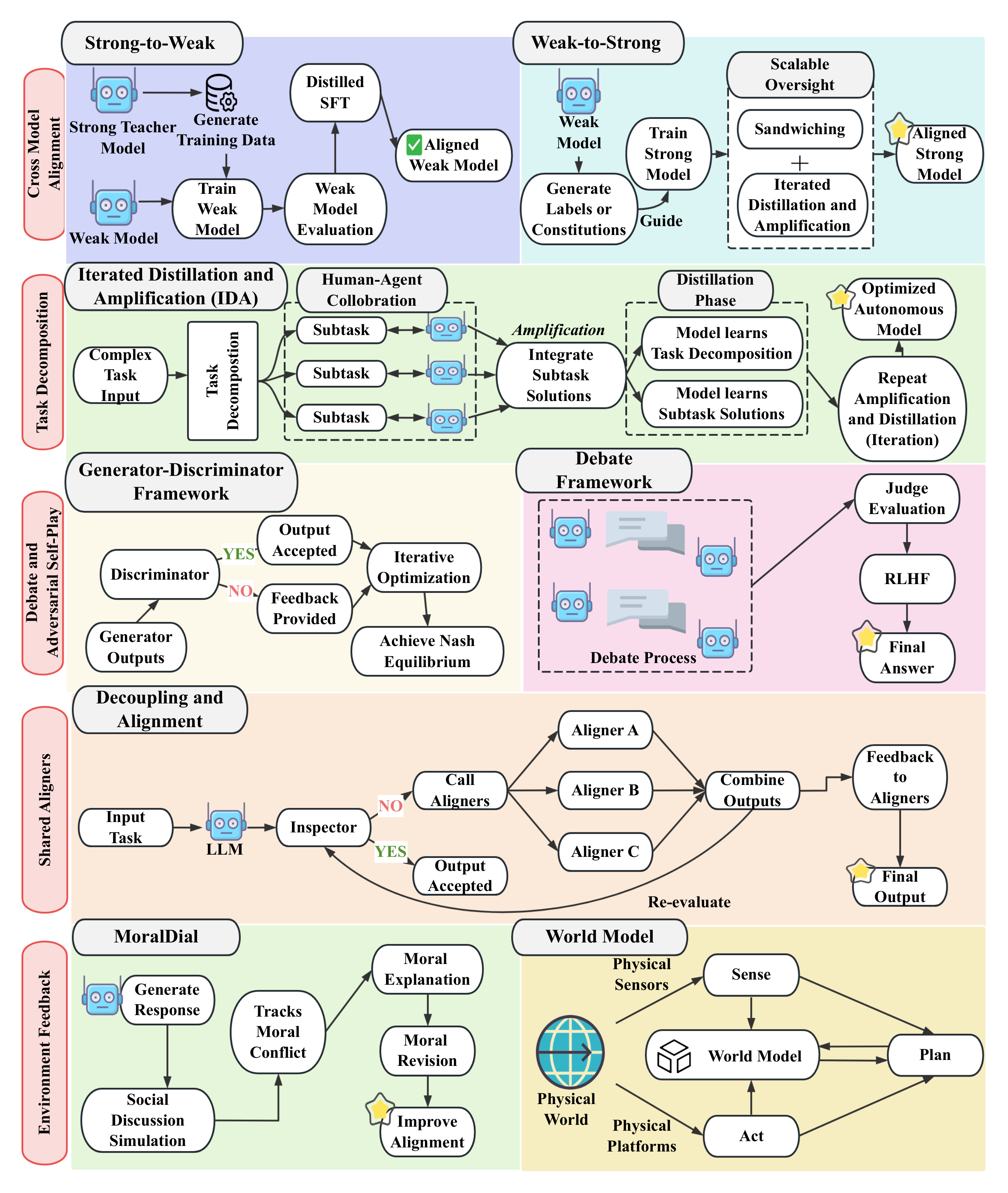}
    \end{adjustbox}
    \caption{Taxonomy and Detailed visualization of Agent-to-Agent Agreement Techniques. }
    \label{Debate}
    \vspace{-12pt}
\end{figure}


\newpage

\end{document}